\documentclass[conference]{IEEEtran}
\IEEEoverridecommandlockouts
\usepackage{cite}
\usepackage{amsmath,amssymb,amsfonts}
\usepackage{algorithmic}
\usepackage{graphicx}
\usepackage{url}
\usepackage{hyperref}
\usepackage{textcomp}
\usepackage{comment}
\usepackage{xcolor}
\usepackage[top=0.8in, bottom=1.05in, left=0.8in, right=0.8in, includefoot]{geometry}
\def\BibTeX{{\rm B\kern-.05em{\sc i\kern-.025em b}\kern-.08em
    T\kern-.1667em\lower.7ex\hbox{E}\kern-.125emX}}
\begin{document}

\title{An Incentive-Aligned Trust-Centric Coordination Layer for Heterogeneous Blockchain Interoperability}

\author{
  \IEEEauthorblockN{
    Saeed Moradi\IEEEauthorrefmark{1},
    Sara Rouhani\IEEEauthorrefmark{2}
  }
  \IEEEauthorblockA{\IEEEauthorrefmark{1}Department of Computer Science, University of Manitoba, Winnipeg, Canada \\
  \IEEEauthorrefmark{2}Department of Computer Science, University of Calgary, Calgary, Canada \\
  Email: moradis1@myumanitoba.ca, sara.rouhani@ucalgary.ca}
}

\maketitle

\begin{abstract}
The interoperability between permissioned and public blockchains is hampered by asymmetric privacy, trust, and governance assumptions. 
We propose a coordination architecture that decouples cross-domain trust management and privacy enforcement from underlying blockchain protocols, requiring no protocol-level modifications. The framework integrates cryptographic unlinkability mechanisms with incentive-compatible relay selection, combining stake-weighted participation and temporally accumulated reputation to align economic commitment with sustained correctness. We formalize the adversarial model and derive security and privacy properties for cross-domain coordination, demonstrating robustness against Sybil infiltration, collusion, and strategic stake concentration. Simulation-based analysis characterizes key design trade-offs between quorum thresholds, reputation memory, and anonymity parameters, indicating that privacy-preserving coordination can be achieved without degrading responsiveness or incentive stability when parameters are properly tuned.
\end{abstract}


\section{Introduction}
\label{sec:introduction}

Blockchain ecosystems are becoming increasingly interconnected; however, interoperability between permissioned and permissionless networks remains a fundamental challenge. Existing cross-chain solutions such as Polkadot \cite{burdges2020overview}, Cosmos \cite{cosmos2019}, Chainlink \cite{breidenbach2021chainlink}, and Axelar \cite{axelar2021} primarily facilitate interoperability across public (permissionless) blockchains under relatively homogeneous assumptions about trust and participation. These systems typically rely on open validator sets, global state visibility, and economically incentivized security models, which do not readily translate to enterprise environments. In contrast, permissioned networks impose restrictive access control, limited data visibility, and governance-driven trust models, creating a fundamental mismatch that limits the applicability of existing interoperability approaches \cite{hassan2026sok}.

In practice, many emerging applications require cross-domain coordination across heterogeneous settings, where sensitive data originates in permissioned systems but must interact with public infrastructures. Examples include enterprise data sharing with public auditability, cross-chain financial settlement between consortium and DeFi systems, and hybrid governance workflows spanning regulated and open networks.

This gap is not only technical but structural. Permissionless blockchains are designed around open participation, public verifiability, and economic consensus, whereas permissioned systems enforce authenticated membership, governance policies, and strict privacy requirements. Bridging these fundamentally different models introduces asymmetric trust, privacy, and governance assumptions that fall outside the design scope of existing interoperability solutions. As a result, current frameworks either restrict themselves to homogeneous environments or implicitly weaken enterprise privacy and accountability guarantees when interacting with public blockchains.

Beyond privacy, interoperability introduces a trust gap. Cross-chain communication depends on intermediate entities (e.g., relays or bridges) that are not inherently trusted by either domain. Purely stake-based mechanisms enable capital-driven attacks, while reputation-only approaches remain vulnerable to long-term manipulation \cite{wang2023reputation}. Designing incentive-compatible and adversary-resilient coordination mechanisms therefore remains an open challenge.

In this work, we formulate interoperability as a \emph{cross-network coordination problem under asymmetric trust and privacy constraints}, rather than a simple bridging task. Building on this perspective, we propose a coordination layer that operates independently of underlying blockchain protocols and achieves the following goals:

\begin{itemize}
    \item \textbf{Privacy-preserving interoperability:} We design a cross-chain coordination framework that integrates cryptographic primitives, including ring signatures, blind signatures, and source-anonymization protocols, to ensure unlinkability and protect metadata originating from permissioned domains.
    
    \item \textbf{Incentive-aligned trust management:} We introduce a hybrid stake--reputation mechanism in which relay nodes are evaluated based on both economic commitment and sustained behavioral performance, enabling robust and manipulation-resistant participation.
    
    \item \textbf{Adversary-resilient coordination:} We formalize the threat model and design the system to mitigate key adversarial strategies, including Sybil attacks, collusion, and replay attacks, through a combination of weighted quorum rules and dynamic trust updates.
    
    \item \textbf{Protocol-agnostic deployability:} We decouple cross-chain coordination from underlying blockchain protocols, enabling interoperability without requiring protocol-level modifications and preserving native governance and consensus assumptions.
    
    \item \textbf{Comprehensive evaluation and trade-off analysis:} We implement a prototype and conduct experiments to analyze trade-offs between privacy, security, latency, and incentive dynamics. The evaluation characterizes the impact of key parameters (e.g., quorum threshold, reputation memory, and anonymity settings) and shows that the framework achieves privacy-preserving coordination without significant loss in responsiveness or incentive stability.
\end{itemize}

To realize these contributions, we design and develop a relay-based coordination framework that integrates cryptographic anonymity mechanisms with incentive-driven trust management. The framework decouples cross-chain coordination from native blockchain logic, enabling secure interaction across heterogeneous environments without compromising their underlying assumptions.

\section{Related Work}

Blockchain interoperability has gained increasing attention across applications such as healthcare~\cite{puneeth2024cross,sonkamble2024secure}, IoT and smart grids~\cite{rahman2022blockchain,singh2025blockchain}, and peer review~\cite{tenorio2021decentralizing}. However, many solutions primarily target permissionless ecosystems, leaving challenges in heterogeneous trust, privacy, and governance insufficiently addressed.

\subsection{Permissionless Interoperability Frameworks}

\textbf{Polkadot} connects heterogeneous parachains through shared relay-chain security~\cite{wood_polkadot}, while \textbf{Cosmos} enables cross-chain communication through IBC~\cite{cosmos2019}, requiring compatible protocols and light-client support. \textbf{Axelar} provides validator-based cross-chain communication~\cite{axelar2021}. Although these frameworks facilitate interoperability, their security and governance models introduce integration challenges for enterprise environments with stricter privacy and accountability requirements.

\subsection{Permissioned Interoperability Frameworks}

\textbf{Hyperledger Cacti} provides modular connectors for heterogeneous blockchains, including Fabric, Corda, Quorum, and public blockchain platforms~\cite{androulaki2018hyperledger,brown2016corda,quorum,hyperledgercacti2025}. However, interoperability connectivity alone does not fully address cross-domain privacy and incentive alignment. Oracle-based approaches~\cite{alhussayen2024blockchain} and publish--subscribe architectures~\cite{ghaemi2021pub} facilitate cross-chain communication but face limitations in scalability, adversarial robustness, or broader transaction functionality~\cite{khan2021towards}. Diallo et al.~\cite{diallo2024agnostic} integrate zero-knowledge proofs and incentive-aligned gateways, primarily targeting private ecosystems.

\subsection{Permissioned--Permissionless Interoperability}

Mitani et al.~\cite{mitani2020traceability} investigate privacy-preserving interoperability using sidechains. However, cross-domain incentive alignment, metadata leakage, and adversarial coordination remain insufficiently addressed in existing heterogeneous interoperability solutions.

\subsection{Privacy-Preserving Cross-Chain Techniques}

CoinShuffle~\cite{ruffing2014coinshuffle} and MixCoin~\cite{bonneau2014mixcoin} enhance transaction unlinkability, while Xclaim~\cite{zamyatin2019xclaim} focuses on secure cross-chain asset transfers. P2C2T~\cite{han2025p2c2t} provides atomicity and unlinkability but does not explicitly address intermediary collusion under heterogeneous governance. Our approach combines ring signatures, blind signatures, and source-anonymization protocols to protect enterprise-originated cross-chain interactions.

\subsection{Incentive Mechanisms}

Incentive mechanisms are essential for cross-chain security~\cite{huang2019survey}. Ye et al.~\cite{ye2025cross} propose PoS-based trust aggregation for IoT sidechains, while game-theoretic approaches such as GTI~\cite{li2023game} and reputation-based consensus mechanisms such as BRBC~\cite{deoliveira2020brbc} address incentive alignment and trust. However, these approaches do not jointly address privacy preservation, dynamic relay trust, and asymmetric governance across permissioned and permissionless blockchains.

\subsection{Discussion and Research Gap}

Despite extensive research on blockchain interoperability, the integration of permissioned and permissionless systems remains insufficiently addressed. Existing frameworks either assume homogeneous trust within permissionless ecosystems or restrict interoperability to closed, enterprise-only environments. Moreover, many approaches introduce intermediary bridge components without systematically addressing privacy leakage or incentive alignment, leaving them vulnerable to bridge-targeted attacks \cite{zhao2023comprehensive}.

The asymmetry between permissioned and permissionless blockchains presents both challenges and opportunities. Permissionless networks typically experience lower throughput due to consensus latency, whereas permissioned systems can achieve higher performance under authenticated participation \cite{esmaili2025performance, amiri2019caper}. Conversely, small consortium-based permissioned networks may exhibit weaker security guarantees compared to large validator sets in public blockchains \cite{nasir2024securing}. These structural differences suggest that cross-domain integration could enable complementary security and performance benefits, provided privacy constraints and governance boundaries are preserved.

However, existing solutions rarely provide a unified treatment of privacy preservation, trust management, and incentive compatibility under asymmetric governance assumptions. To address this gap, we propose a generic coordination framework that connects permissioned and permissionless blockchains under minimal assumptions. Inspired by \cite{Moradi2025PrivacyPreserving}, the design introduces a relay-based bridging component supported by smart contracts and privacy-preserving cryptographic primitives, including ring and blind signatures. A hybrid reputation–stake incentive mechanism further aligns participation with sustained honest behavior, mitigating Sybil, collusion, and strategic manipulation in heterogeneous environments.

\section{System Design}

\subsection{System Overview}

The proposed architecture introduces an intermediate coordination layer that enables secure interaction between a permissioned and a permissionless blockchain without modifying their native protocols. The system consists of three main roles: (i) \emph{watchers}, which observe events on the permissioned network, (ii) \emph{relays}, which anonymize and forward messages, and (iii) \emph{smart contracts}, which verify, finalize, and evaluate relay behavior.

Cross-chain interaction proceeds in three phases: (1) event observation and anonymization, (2) relay-based forwarding and quorum agreement, and (3) on-chain verification and trust update. The system assumes a partially honest majority of relay weight and bounded network delays.

\subsection{System Components}

\textbf{Watcher Nodes:} Monitor the permissioned blockchain and emit encrypted event tokens corresponding to relevant transactions. To mitigate potential collusion among watchers, the system relies on redundancy and cross-verification across multiple independent observers, making it difficult for a small subset to bias event propagation without detection.

\textbf{Relay Nodes:} Retrieve anonymized tokens, reconstruct payloads, and forward them to the destination blockchain.

\textbf{Bridge Contract:} Collects relay submissions, enforces quorum agreement, and finalizes accepted messages.

\textbf{Trust Contract:} Maintains relay trust scores and stake metadata, updates reputation, and distributes rewards and penalties.

\textbf{Off-chain Auditor:} Computes computationally intensive statistics and submits verifiable summaries to the Trust Contract. While the current design assumes a logically trusted auditor for efficiency, the outputs are cryptographically verifiable, and the role can be extended to a decentralized committee or replicated across multiple independent entities to reduce trust assumptions.

\subsection{Transaction Flow}

A cross-chain interaction proceeds as follows:

\begin{enumerate}
    \item A transaction on the permissioned blockchain emits an event with hash $h = \text{Hash}(m \parallel \text{nonce})$.
    \item Watchers observe the event, encrypt it, and propagate it using an anonymization protocol.
    \item Relay nodes retrieve anonymized tokens, reconstruct the message, and generate cryptographic proofs.
    \item Relays submit candidate message hashes to the Bridge Contract.
    \item A message is finalized once a quorum of relays agrees on a consistent hash.
    \item Relays submit proofs, which are used to update trust scores and distribute rewards.
\end{enumerate}

\subsection{Privacy Mechanisms}

To preserve transaction privacy across domains, the framework integrates three complementary mechanisms.

\textbf{Blind signatures} enable authorization without revealing message content or relay identity, ensuring unlinkability between approval and forwarding.

\textbf{Ring signatures} provide anonymity within a set of relays, preventing identification of the specific relay endorsing a message.

\textbf{Source-anonymization protocols} Obfuscate network-level metadata such as IP address and timing.

Together, these mechanisms ensure unlinkability across both cryptographic and network layers.

\subsection{Trust and Incentive Model}

The Trust Contract updates relay trust scores based on observable behavior derived from on-chain events, relay proofs, and watcher attestations. Metrics include participation rate $p_i^{(t)}$, success rate $s_i^{(t)}$, delay score $d_i^{(t)}$, validated challenge rate $c_i^{(t)}$, and collusion indicator $\kappa_i^{(t)}$.

The per-epoch performance score is defined as:

\begin{equation}
R_i^{(t)} = w_p p_i^{(t)} + w_s s_i^{(t)} + w_d d_i^{(t)} - w_c c_i^{(t)} - w_\kappa \kappa_i^{(t)}.
\end{equation}

This score is normalized via:

\begin{equation}
\sigma(R) =
\frac{R + (w_c + w_\kappa)}{1 + (w_c + w_\kappa)}.
\end{equation}

Trust is updated using an exponentially weighted moving average (EWMA):

\begin{equation}
\label{eq:ewma}
T_i^{(t+1)} = \text{clip}_{[0,1]}\left(\alpha T_i^{(t)} + (1 - \alpha)\sigma(R_i^{(t)})\right).
\end{equation}

To mitigate wealth concentration, stake is concavely normalized:

\begin{equation}
S_i^e =
\frac{\log(1 + S_i)}{\log(1 + S_{\max})}.
\label{eq:stakenorm}
\end{equation}

The effective weight of each relay is defined as:

\begin{equation}
w_i^{(t)} =
T_i^{(t)} \left(\omega_R + (1 - \omega_R) S_i^e \right),
\label{eq:weight}
\end{equation}

which determines both quorum participation and reward allocation.

A candidate message $h^*$ is accepted if:

\begin{equation}
\sum_{i : h_i = h^*} w_i^{(t)} \geq \tau \sum_k w_k^{(t)}.
\label{eq:quorum}
\end{equation}

Rewards are distributed proportionally:

\begin{equation}
\text{Reward}_i^{(t)} =
R_{\text{pool}}^{(t)} \frac{w_i^{(t)}}{\sum_k w_k^{(t)} + \epsilon}.
\label{eq:reward}
\end{equation}

\subsection{Collusion Detection}

To mitigate coordinated manipulation, a collusion indicator $\kappa_i^{(t)}$ is computed based on agreement statistics among relays. The weighted agreement score is:

\begin{equation}
A_i^{(t)} =
\frac{\sum_{j \ne i} w_j^{(t)} \sum_m \mathbf{1}_{i,j}^{(m)}}
{\sum_{j \ne i} w_j^{(t)} \cdot N^{(t)}}.
\end{equation}

The collusion indicator is defined as:

\begin{equation}
\kappa_i^{(t)} =
\text{clip}_{[0,1]}
\left(
\frac{A_i^{(t)} - \mu_A^{(t)}}{\kappa_\sigma \sigma_A^{(t)}}
\right).
\label{eq:kappa}
\end{equation}

Relays with anomalously high agreement scores are signaled through reduced trust and influence in subsequent epochs.

\section{Threat Model}
To evaluate the privacy and security of our system, we adopt the STRIDE threat model \cite{shostack2014threat}, followed by sybil and collusion attacks which are typical of decentralized environments.

\
\paragraph{Spoofing}
Spoofing involves impersonating legitimate relays to gain unauthorized access~\cite{shostack2014threat}. Our system requires cryptographically registered identities, collateral deposits, and on-chain verification of relay signatures and proofs. These mechanisms prevent impersonation without access to the corresponding private keys.

\paragraph{Tampering}
Tampering involves unauthorized modification of data~\cite{shostack2014threat}. Source-chain signatures and proofs are verified by the Bridge Contract, rejecting altered submissions. Watcher nodes can additionally challenge invalid messages, triggering penalties through slashing.

\paragraph{Repudiation}
Repudiation occurs when actors deny responsibility for their actions~\cite{shostack2014threat}. Relay submissions, challenges, and adjudications are recorded as auditable on-chain events, providing verifiable evidence of recorded actions and supporting accountability.

\paragraph{Information Disclosure}
Information disclosure pertains to unauthorized leakage of sensitive data \cite{shostack2014threat}. In bridging between permissioned and permissionless networks, sender anonymity and confidentiality are critical. To address this, the system employs ring signatures, blind signatures, and source-anonymization protocols. Relays and watchers only observe blinded or encrypted payloads, preventing them from inferring the identity of the sender or the contents of confidential transactions.

\paragraph{Denial of Service (DoS)}
DoS attacks aim to disrupt system availability by preventing honest relays from participating or by flooding the bridge \cite{shostack2014threat}. The quorum mechanism ensures liveness so long as a sufficient fraction of effective weight behaves honestly, while the stake requirement discourages large-scale Sybil-based spam. Additionally, the trust update mechanism penalizes relays that repeatedly fail to participate or that introduce excessive delay. These measures mitigate protocol-level DoS threats. However, network-level flooding attacks remain outside the scope of the proposed architecture. Mitigating DoS attacks becomes more challenging especially when anonymization protocols such as Dandelion++ are adopted \cite{cypherpunk_dandelion}.

\paragraph{Elevation of Privilege}
Elevation of privilege occurs when adversaries gain unauthorized control or disproportionate influence~\cite{shostack2014threat}. Our design mitigates this threat through reputation-weighted, concavely normalized stake, collusion detection, and slashing mechanisms that discourage malicious behavior and limit excessive influence.

\subsection{Sybil Resistance}
Sybil attacks allow adversaries to amplify their influence through multiple identities. Our framework raises the cost of such attacks by requiring a minimum collateral deposit \(S_{\min}>0\) per relay and combining concavely normalized stake (Eq.~\eqref{eq:stakenorm}) with reputation-gated weighting (Eq.~\eqref{eq:weight}). New identities therefore require both collateral and accumulated trust to gain substantial influence. While concave normalization limits the advantage of concentrated stake, it does not inherently prevent stake-splitting attacks. Together, these mechanisms increase the economic and temporal costs of Sybil attacks without guaranteeing complete resistance.


\subsection{Collusion detection and response}
Even if Sybil inflation is infeasible, adversaries may attempt to collude by coordinating submissions among a small group of relays. 
Such behavior can bias quorum formation or suppress diversity of opinions. 
To mitigate this, we introduce a collusion indicator \(\kappa_i^{(t)}\) (Eq.~\eqref{eq:kappa}), which signals whether a relay’s agreement rate with a small subset of peers is statistically anomalous relative to the population. Relays with unusually high pairwise agreement scores are flagged as potential colluders. Upon detection, the Trust Contract enforces defensive measures: (i) challenge rewards are dynamically increased to incentivize external verification in the suspicious epoch, (ii) the trust scores $T_i^{(t)}$ of implicated relays are reduced via the exponential moving average update (Eq.~\eqref{eq:ewma}), and (iii) the Bridge Contract can require additional watcher attestations before finalizing any message supported primarily by the colluding set. These responses jointly reduce the effective weight of colluders, expose them to greater challenge risk, and limit their capacity to bias quorum outcomes.



\subsection{Source Anonymization Security Analysis}

To mitigate \textit{de-anonymization attacks}, our system leverages anonymization protocols that obfuscate the origin of transactions by routing messages through randomly selected relays in a forwarding path. Such attacks, as described in \cite{zhang2020txspector}, exploit the susceptibility of the initial broadcasting node, enabling adversaries to infer transaction origins if enough forwarding relays are compromised.

\subsubsection{Adversarial Model}
Let the forwarding path consist of $f$ relays, drawn from a set of $n$ total relays. In our design, relays are not chosen uniformly at random but are instead selected with probability proportional to their trust-weighted score $w_i^{(t)}$ (Eq.~\ref{eq:weight}). Thus, the effective probability of a malicious relay being chosen is
\[
  p_\mathsf{adv} = \sum_{i \in \mathcal{A}} \frac{w_i^{(t)}}{\sum_{k=1}^n w_k^{(t)}},
\]
where $\mathcal{A}$ denotes the set of relays controlled by adversaries.  In contrast to uniform selection, this approach biases exposure towards high-trust relays, which are limited in quantity yet strategically appealing targets.

\section{Evaluation}

We evaluate the proposed architecture using a prototype implementation that integrates a permissioned blockchain (Hyperledger Fabric \cite{monrat2020performance, polge2021permissioned}) and a permissionless blockchain (Substrate contract node \cite{paritytech_substrate_contracts_node}). The evaluation is based on an implemented prototype integrating Hyperledger Fabric and Substrate, where the coordination layer, relay behavior, and incentive mechanisms are modeled to analyze system-level properties under varying conditions.

The implementation and experimental scripts are available in an anonymized repository \footnote{\url{https://anonymous.4open.science/r/thesis_project-1BAB/README.md}}

\begin{figure}[t]
    \centering
    \includegraphics[width=\columnwidth]{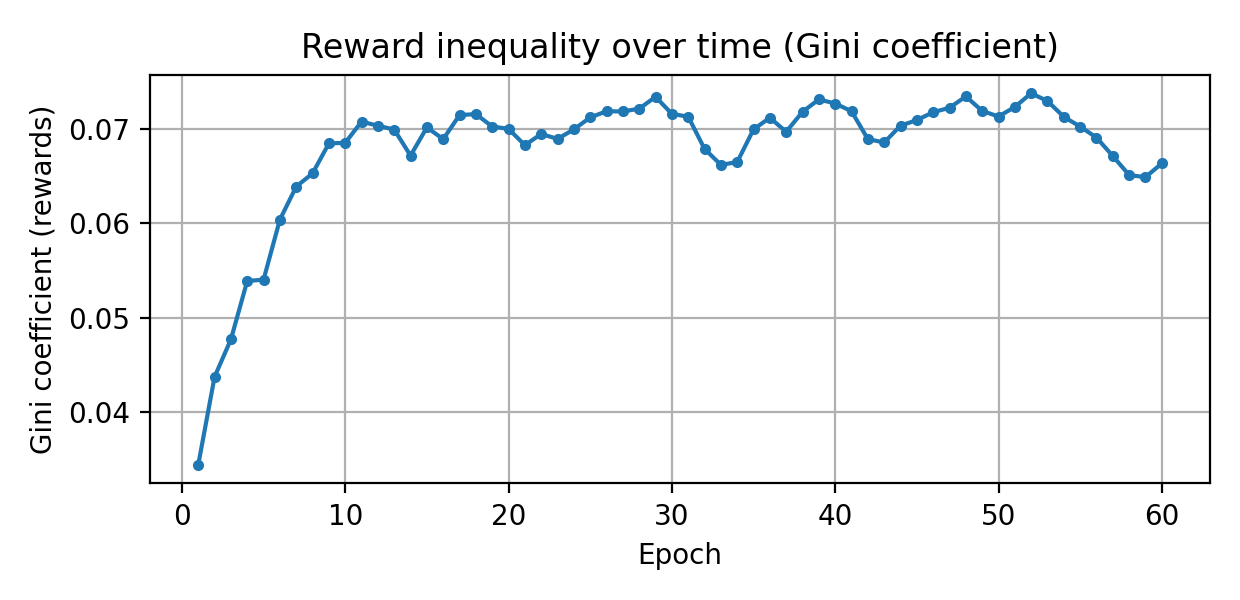}
    \caption{Reward inequality over time, measured using the Gini coefficient per epoch. }
    \label{fig:gini-time}
\end{figure}

\subsection{Reward Inequality Over Time (Gini Coefficient)}
\label{subsec:gini-time}

Figure~\ref{fig:gini-time} shows the epoch-by-epoch Gini coefficient computed from per-epoch reward allocations. The curve increases sharply during the initial reputation-formation phase and then stabilizes with modest fluctuations.  

\textbf{Interpretation:} The early rise reflects stochastic advantages that are amplified through the EWMA-based trust update in combination with multiplicative stake–reputation weighting. Once EWMA scores converge, the Gini stabilizes: long-term concentration remains modest because reputation dynamics and challenge penalties limit runaway dominance.  

\textbf{Implications:} Larger values of \(\alpha\) slow adaptation and prolong inequality, whereas smaller values reduce long-term concentration but introduce higher volatility. This temporal profile suggests that \(\alpha\) must be tuned to balance stability and responsiveness. If necessary, onboarding protections (e.g., temporary weight boosts for new relays) could be introduced to mitigate early-phase inequality.

\begin{figure}[t]
    \centering
    \includegraphics[width=\columnwidth]{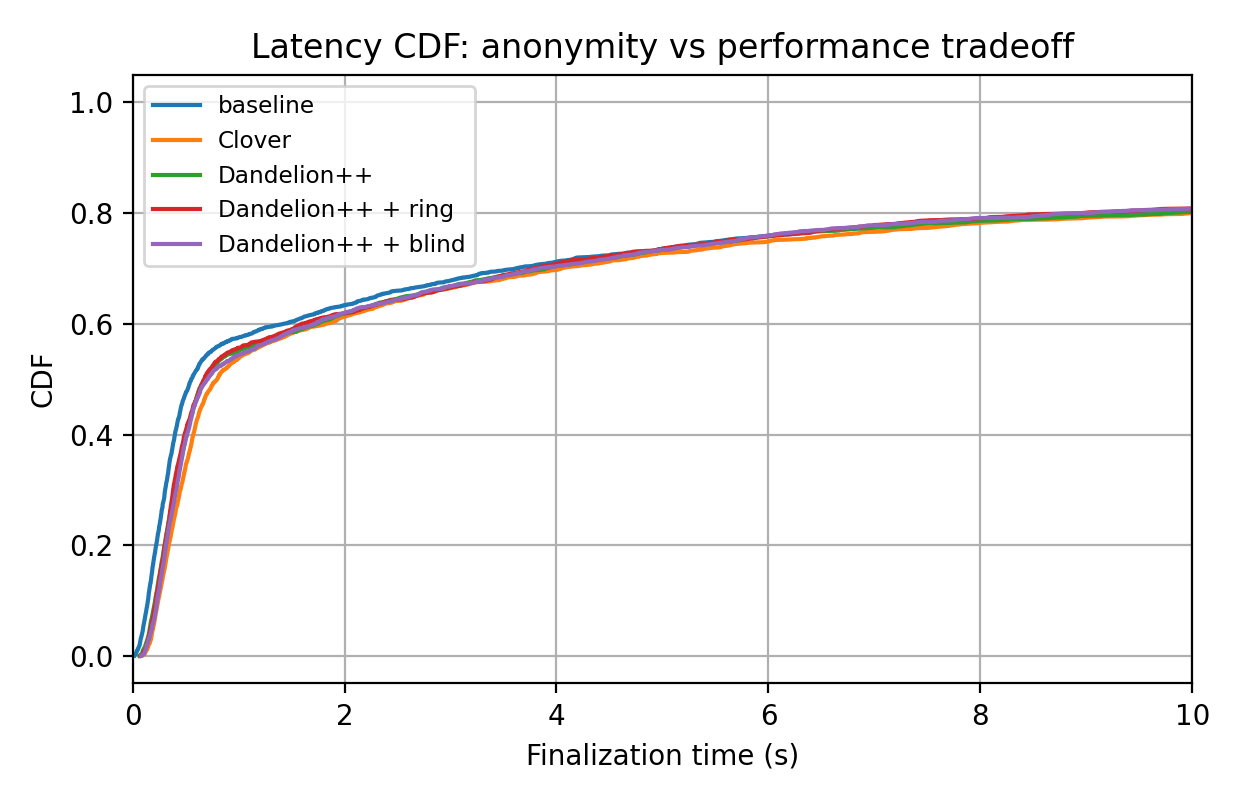}
    \caption{CDF of end-to-end message finalization latency for different anonymization protocols and cryptographic configurations.}
    \label{fig:latency-cdf}
\end{figure}

\subsection{Latency CDF (Anonymity vs.\ Performance Trade-off)}
\label{subsec:latency-cdf}

Figure~\ref{fig:latency-cdf} presents the empirical CDF of finalization latency across multiple configurations (baseline, Clover, Dandelion++, and cryptographic overheads from ring and blind signatures). While the curves remain closely clustered, modest shifts in both median and tail latencies are observed due to anonymization and cryptographic processing.  

\textbf{Interpretation:} Anonymity mechanisms increase message latency through additional hops, dummy traffic, and cryptographic operations. The most pronounced effects appear in the distribution tail, where extended paths and sporadic drops by colluding relays lead to longer completion times. Median latency increases only slightly, indicating that overhead is concentrated in adverse cases rather than in typical performance.  

\textbf{Implications:} To satisfy stringent low-latency service-level agreements (SLAs), anonymization parameters (e.g., path length, stem duration) must be tuned carefully, or applied selectively to privacy-sensitive messages. Furthermore, retries and timeouts should be incorporated to contain the impact of adversarial delays, with tail latency monitored as a critical performance metric.
This result indicates that privacy-preserving cross-network coordination can be achieved without sacrificing system responsiveness, provided that relay participation thresholds are tuned.

\begin{figure*}[t]
    \centering
    \includegraphics[width=0.95\textwidth]{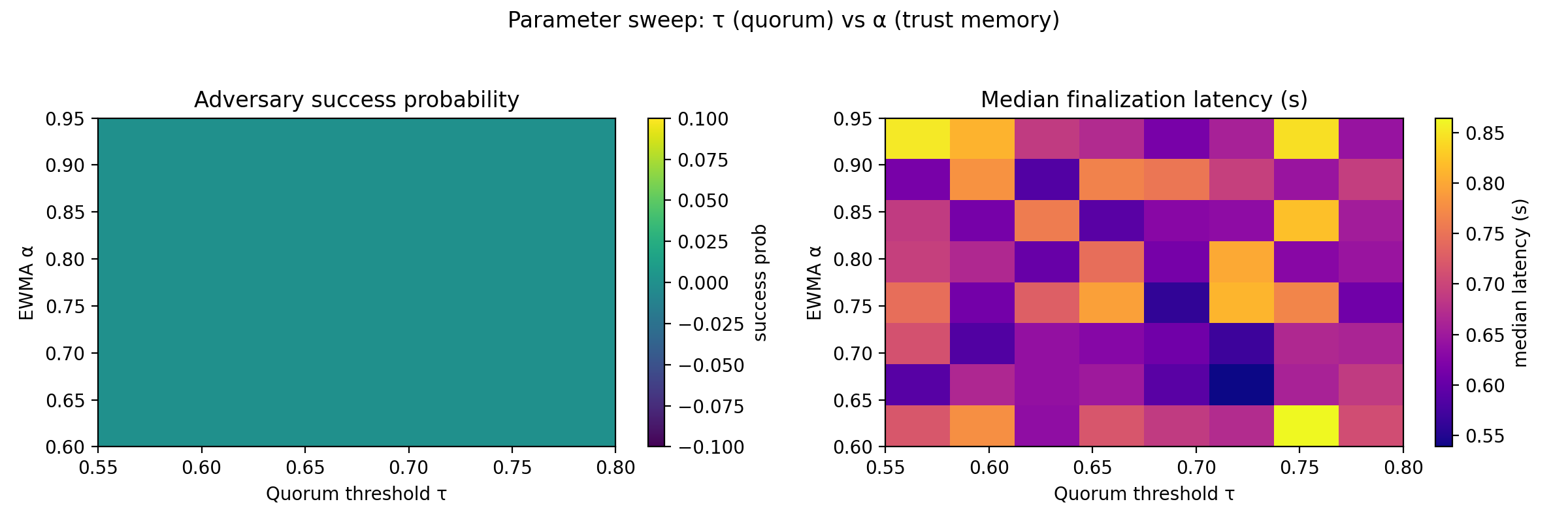}
    \caption{Parameter sweep results for quorum threshold $\tau$ and EWMA memory $\alpha$. Left: estimated adversary success probability under the tested model. Right: median finalization latency. }
    \label{fig:param-sweep}
\end{figure*}

\subsection{Parameter Sweep: Quorum Threshold \texorpdfstring{$\tau$}{tau} vs.\ EWMA Memory \texorpdfstring{$\alpha$}{alpha}}
Figure~\ref{fig:param-sweep} demonstrates two heatmaps over the \((\alpha,\tau)\) grid: (left) estimated adversary success probability under the experiment’s proxy adversary model; (right) median finalization latency. In this run, adversary success remained uniformly low across the sampled grid, while latency varied significantly.  

\textbf{Interpretation:} The results confirm the expected safety–liveness trade-off. Higher values of \(\tau\) reduce adversary success probability by raising the effective-weight threshold required for finalization, but also increase latency since more endorsements must be accumulated. The parameter \(\alpha\) regulates the responsiveness of reputation: larger values slow adaptation, which can exacerbate latency when high-weight relays are temporarily unavailable.  

\textbf{Implications:} For conservative deployments, \(\tau\) should be chosen near the upper end of the recommended range, with \(\alpha\) set in a moderate interval (e.g., \(\alpha \in [0.75,0.9]\)). For latency-sensitive environments, reducing \(\tau\) or shortening anonymization path lengths can lower completion times, though at the cost of weaker adversarial resistance.
More broadly, this result highlights that security in asymmetric cross-network interoperability is not achieved through static conservatism, but through balanced co-design of quorum and reputation dynamics that jointly control adversarial influence and system responsiveness.

\begin{figure}[t]
    \centering
    \includegraphics[width=\columnwidth]{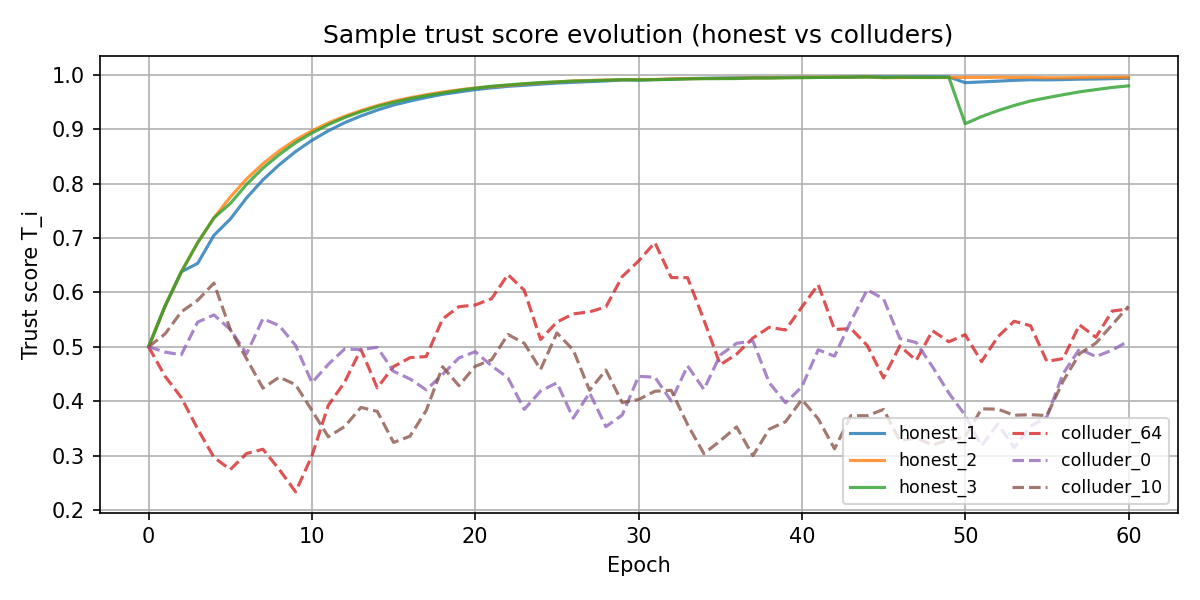}
    \caption{Trust-score trajectories for representative honest and malicious relays.}
    \label{fig:trust-time-series}
\end{figure}

\subsection{Trust Score Time Series}
Figure~\ref{fig:trust-time-series} illustrates sample trust-score trajectories for a subset of relays over the full simulation, including representative honest nodes and several identified colluders (malicious relays). Trust scores evolve through the EWMA update (Eq.~\eqref{eq:ewma}), incorporating participation, correctness, delay, and penalties from challenges or collusion.  

\textbf{Interpretation:} Honest relays exhibit stable or gradually increasing trust scores due to consistent participation and accurate forwarding, whereas colluders show declining trajectories driven by message drops, elevated delays, and frequent challenges. The divergence between honest and malicious nodes highlights the EWMA’s ability to integrate both recent behavior and historical reputation, progressively demoting misbehaving relays.  

\textbf{Implications:} These dynamics confirm that reputation functions as both informative and actionable. Deteriorating behavior is penalized quickly via the \((1-\alpha)\,\sigma(R)\) term, while sustained honest behavior maintains or restores influence. Consequently, $T_i^{(t)}$ in Eq.~\eqref{eq:ewma} provides a reliable basis for regulating stake influence and guiding quorum decisions.
The observed trust score dynamics highlight that reputation must be treated as a temporal control signal rather than a static metric, enabling resilient coordination under asymmetric trust and intermittent adversarial behavior.

\subsection{Reward histogram (final epoch)}

\begin{figure}[t]
    \centering
    \includegraphics[width=\columnwidth]{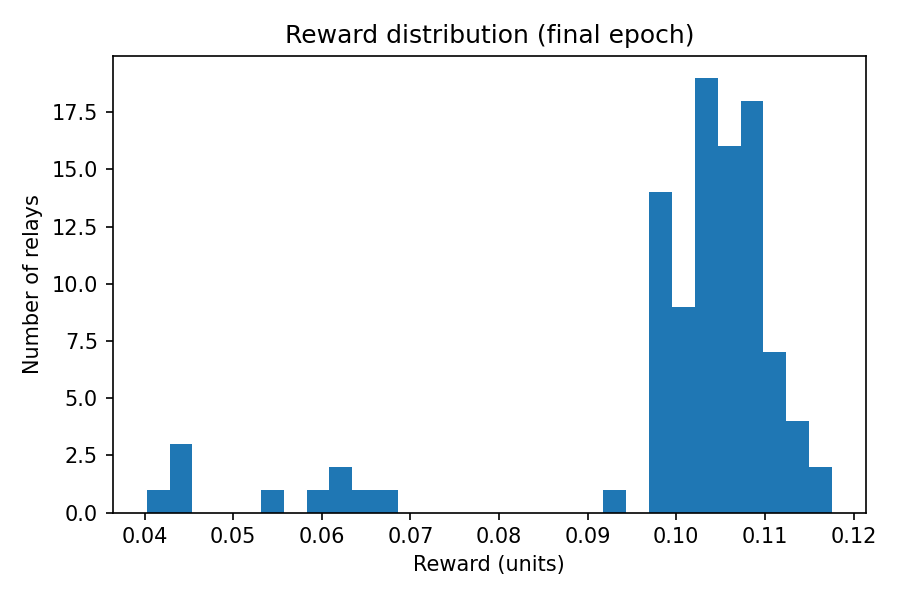}
    \caption{Histogram of cumulative rewards per node}
    \label{fig:reward-histogram}
\end{figure}

Figure~\ref{fig:reward-histogram} shows the per-node reward distribution at the final epoch of the simulation. The distribution reflects the combined effects of effective weight \(w_i^{(t)}\), the reward pool, and the epochal allocation rule (Eq.~\eqref{eq:reward}).  

\textbf{Interpretation:} Rewards concentrate among nodes that have consistently accumulated both stake and reputation, i.e., those with high trust scores $T_i^{(t)}$ and nontrivial normalized stake \(\widetilde S_i\) as defined in Eq.~\eqref{eq:stakenorm}. The histogram, which is not significantly heavy-tailed, indicates that reputation gating prevents a small set of high-stake but low-reputation relays from capturing the majority of rewards.  

\textbf{Implications:} These results support the claim that the hybrid stake–reputation mechanism yields incentive-compatible outcomes: reliable relays with sustained good performance are rewarded, while malicious or inactive relays receive negligible payouts. If reward centralization were observed in practice, mitigation strategies such as minimum per-epoch payouts, reward smoothing, or light randomization during early epochs could be introduced to preserve fairness and encourage participation.
In other words, the reward distribution at the final epoch demonstrates that the incentive mechanism effectively aligns long-term economic rewards with sustained honest behavior, while progressively marginalizing unreliable or malicious relays.

\subsection{Attacker Success Probability vs.\ Adversary Stake Fraction}
\begin{figure}[t]
    \centering
    \includegraphics[width=\columnwidth]{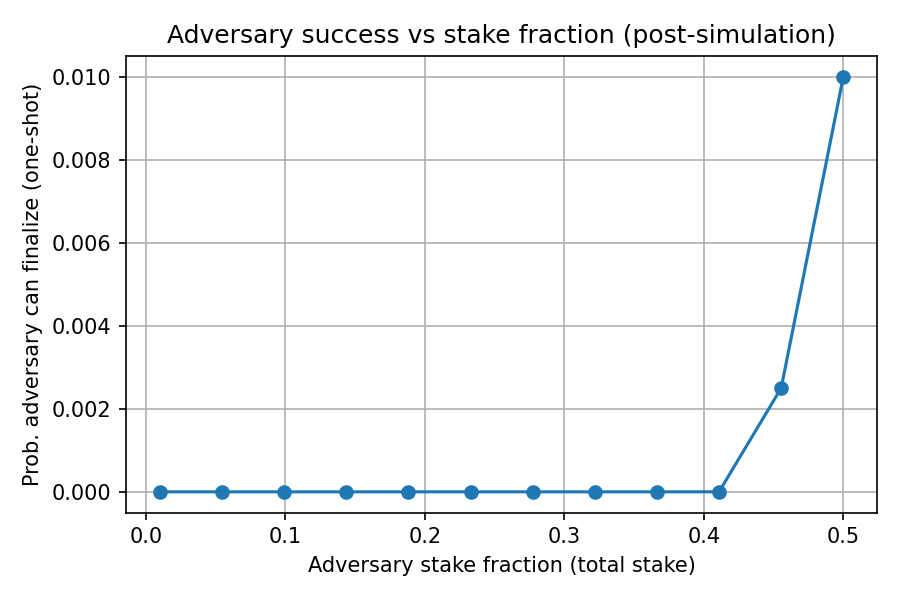}
    \caption{Attacker success probability as a function of total adversarial stake. }
    \label{fig:attacker-success}
\end{figure}

Figure~\ref{fig:attacker-success} reports the empirical probability that an adversary controlling a given fraction of the total stake can assemble sufficient effective weight to satisfy the quorum threshold \(\tau\) and finalize a malicious candidate.

\textbf{Interpretation:} Success probability is negligible for stake fractions below the quorum threshold and rises sharply as the adversary’s stake approaches it. This outcome aligns with the quorum condition (Eq.~\eqref{eq:quorum}) together with multiplicative reputation gating: achieving quorum requires both stake and accumulated reputation, which increases the temporal and economic costs of mounting a successful attack.  

\textbf{Implications:} Under the simulated conditions, the architecture is robust against single-shot stake attacks unless the adversary controls a large share of the stake. These findings support configuring \(\tau\) within the recommended conservative range (\(0.66\)–\(0.75\)). In stronger threat models, resilience can be enhanced by raising \(\tau\) or by extending reputation accumulation windows, thereby further constraining the feasibility of stake-driven takeovers.
This result demonstrates that combining stake-based quorum requirements with reputation gating substantially increases the economic and temporal cost of adversarial takeovers, making single-shot stake attacks impractical in cross-network interoperability settings.

\subsection{Deanonymization Probability vs.\ Ring Size}
\begin{figure}[t]
    \centering
    \includegraphics[width=\columnwidth]{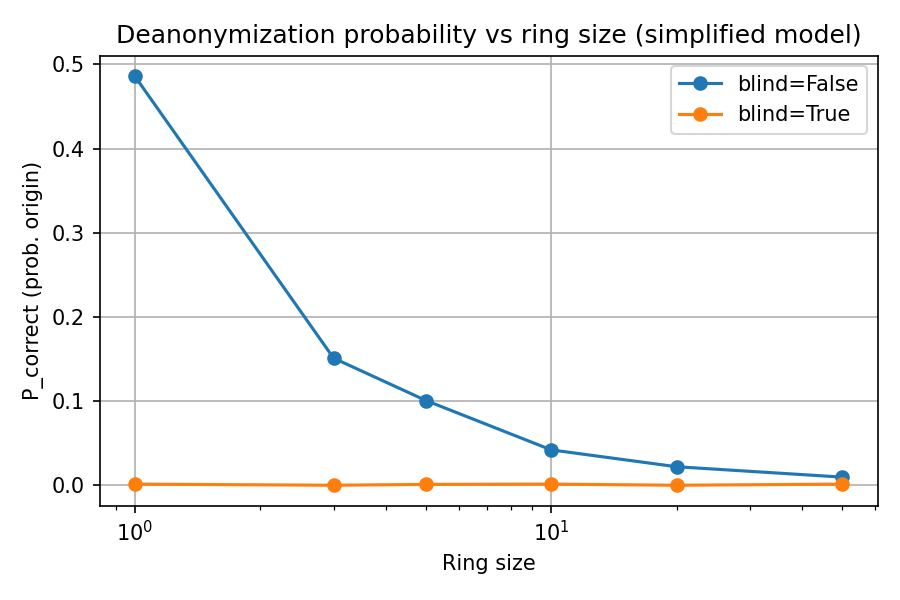}
    \caption{Deanonymization probability versus ring size in the ring signature scheme. }
    \label{fig:deanonymize-ring}
\end{figure}

Figure~\ref{fig:deanonymize-ring} illustrates the probability that an adversary can correctly identify the source of a transaction as a function of the ring size \(r\). For comparison, the plot also includes a curve approximating the effect of blind signatures, which introduce an additional unlinkability gain beyond that of ring signatures alone.  

\textbf{Interpretation:} For ring signatures, deanonymization probability decreases approximately as \(1/r\): larger rings expand the anonymity set, reducing the adversary’s chance of a correct guess. Blind signatures further decouple the signer’s identity from the message, lowering linkage probability in all regimes, with the effect most pronounced for small to moderate ring sizes.  

\textbf{Implications:} From a design perspective, increasing the ring size remains a reliable means of enhancing cryptographic anonymity, while blind signatures provide an additional safeguard against signature-linkage attacks. In practice, the ring size should be selected to balance the target adversarial deanonymization probability \(P_{\text{correct}}\) against processing and latency overheads, ensuring that anonymity goals are met within system performance constraints. 

\subsection{Limitations}
Although the proposed framework substantially reduces the risk of deanonymization under the assumed threat model, it does not fully eliminate metadata leakage in the presence of a global passive adversary with full network visibility. In particular, large-scale traffic correlation across anonymization layers may still enable probabilistic inference of message origins. Rather than pursuing absolute anonymity—which would impose prohibitive latency and coordination overhead—the design intentionally raises the economic and temporal cost of deanonymization beyond practical feasibility through layered cryptographic mechanisms and incentive-aligned relaying. Addressing global traffic analysis and network-wide correlation attacks remains an open challenge for future work.

\section{Conclusion}
This paper presented a privacy-aware, incentive-driven bridge architecture for secure interoperability between permissioned and permissionless blockchains. The design integrates: (i) an anonymization-enabled relay cohort using configurable source-anonymization protocols (Dandelion++ or Clover), (ii) ring and blind signatures to ensure unlinkability and signer anonymity, and (iii) a reputation-aware incentive mechanism implemented via a Trust Contract that combines per-epoch performance metrics with normalized stake to derive effective voting weights and reward allocation. The architecture was evaluated through simulation-based stress testing of scalability and economic robustness, and its privacy, correctness, and security properties were systematically analyzed.

\bibliographystyle{IEEEtran}
\bibliography{references}

\end{document}